\documentclass[a4paper]{iopart} 

\begin{document}

\newcommand{\ds}{\displaystyle}

\thispagestyle{empty}

\title[Singularity Structure, Symmetries and Integrability]{Singularity
Structure, Symmetries and Integrability of Generalized  Fisher Type Nonlinear
Diffusion Equation}

\author{P S Bindu$\dagger$, M Senthilvelan$\ddagger$ and M Lakshmanan$\dagger$}
\address{$\dagger$Centre for Nonlinear Dynamics, Department of Physics,
Bharathidasan University, Tiruchirapalli 620 024, India}
\address{$\ddagger$School of Physics, The University of Sydney, NSW 2006,
Australia}

\date{\today}
\thispagestyle{empty}
\begin{abstract}

In this letter, the integrability aspects of a generalized Fisher  type
equation with modified diffusion in (1+1) and (2+1) dimensions are  studied by
carrying out a singularity structure and symmetry  analysis. It is  shown that
the Painlev\'e property exists only for a special choice  of the parameter
($m=2$). A B\"acklund transformation is shown to  give rise to the linearizing
transformation to the linear heat  equation for this case ($m=2$). A Lie
symmetry analysis also picks  out the same case ($m=2$) as the only system
among this class as  having nontrivial infinite  dimensional Lie algebra of
symmetries and that the similarity  variables and similarity reductions lead in
a natural way to the linearizing  transformation and physically important
classes of solutions  (including known ones in the literature),  thereby giving
a group theoretical understanding of the system.  For nonintegrable cases in
(2+1) dimensions, associated Lie symmetries and similarity reductions are
indicated. 
\end{abstract}

\maketitle

The Fisher type reaction-diffusion equation with quadratic 
nonlinearity and modified diffusion of the form 
\begin{eqnarray} 
u_t - \triangle u-\frac{m}{1-u} ({\nabla u})^2 - u(1-u)=0, 
\end{eqnarray}
where $u(t,x,y)$ is some kinetic variable, $ \triangle $ and  $ \nabla $ are
gradient and Laplacian operators respectively  and the subscript denotes
partial differentiation with respect  to time, is an important physical system
appearing in many areas  of physics and biology [1-4]. The case $m=0$ is the
classical  Fisher equtaion and $m=2$ corresponds to real systems such as 
bacterial colony growth [5]. It has been known for some time that the $m=2$
case can be mapped onto the linear heat equation through a specific
transformation [6]. Recently Brazhnik and Tyson [7]  have  discussed five
interesting classes of travelling wave solutions  and static structures in the
2-dimensional version of the $m=2$  case of eq.(1). In this letter, we wish to
point out that a  Painlev\'e singularity structure analysis [8] picks out the
$m=2$  case for both the (1+1) and (2+1) dimensions as the only system for 
which the partial differential equation  (1)  is free from movable  critical
singular manifolds satisfying the Painlev\'e (P-) property. More interestingly,
we point out that  the B\"acklund transformation deduced from the Laurent
expansion gives rise  to the linearizing transformation in a natural way.
Similarly, a Lie symmetry analysis singles out the $m=2$ case in eq.(1) as the
only system posessing a nontrivial infinite-dimensional Lie algebra of
symmetries  and that the similarity variables and similarity reductions  give
rise to the linearizing  transformation and several physically interesting
solutions,  including the travelling wave solutions, static structures and  so
on known in the literature, in an  automatic way. For the $m \ne 2$ cases, one
obtains restricted classes  of invariant solutions, including propagating
pulses and fronts of special type.

To start with,  we consider the (1+1) dimensional case of eq.(1). Locally
expanding in the neighbourhood of the non-characteristic singular manifold 
$\phi(x,t) = 0,\;\phi_x,\phi_t \ne 0$ in the  form of a Laurent series [8]
\begin{eqnarray}
u=\sum_{j=0}^\infty u_j\phi^{j+p},
\end{eqnarray}
one finds that the possible values of the power of the leading order
term are

\hskip 20 pt (i)\hskip 10 pt
$  p=-2 $,

\hskip 20 pt (ii)\hskip 10 pt
$p=\displaystyle{\frac{1}{1-m}},\quad m \ne 1,$

\hskip 20 pt (iii) \hskip 10 pt$p = 0.$ \\
One can easily check that for all these three leading order cases,
only for the value $m=2$ the solution is free from movable critical
singular manifolds. One finds that in the case $p=-1$, the leading
 order 
coefficient $u_0$ is an arbitrary function in addition to the 
arbitrary singular manifold $\phi(x,t)$. In the case of the other 
two 
leading orders, for $m=2$ one obtains only one arbitrary function 
without the introduction of movable singular manifolds and so they 
can
be considered as corresponding to special solutions. Thus eq.(1)
in the one-dimensional case is found to satisfy the P-property for
$ m=2$ and 
is expected  to be integrable. For all other choices of $m$, except 
for certain special cases of the constrained singular manifold $\phi_t-\phi_{xx}=0$, the 
solution exhibits the presence of movable critical singular manifold,
and so the system is of non-Painlev\'e type and so nonintegrable. 
Extending this analysis to (2+1) dimensional case also one obtains 
essentially the same conclusion (except for certain cases of the constrained 
manifold $\phi_t-\phi_{xx}-\phi_{yy}=0$).

Now in the Laurent expansion (2) if we cut off the series at 
"constant" level term, that is $j=-p$ for the leading order
$p=1/(1-m)=-1$,  $m=2$, so that 
\begin{eqnarray}
u=\frac{u_0}{\phi}+u_1
\end{eqnarray}
and demand that if $u_1$ is a solution 
of (1) for the case $m=2$ then $u$ is also a solution, one 
essentially has an auto-B\"acklund transformation. Here $u_0$ and $\phi$ satisfy a set of coupled partial differential equations (pdes)
arising from eq.(1) on using the transformation (3). Starting 
from the
trivial solution, $u_1=0$ of eq.(1), one can check that the 
equations 
for $u_0$ and $\phi$ in eq.(3) are consistent for the choice 
$u_0=\phi$, giving rise to the new solution $u=1$ which is indeed
an exact solution of (1). Now taking $u_1=1$ as the new seed 
solution, one can check from equations satisfied by $u_0$ and $\phi$
that
\begin{eqnarray}
u_0=-1,\quad \phi_t-\phi_{xx}-\phi+1=0.
\end{eqnarray}
Defining now $\phi = 1+\chi$, one obtains the linear heat equation
\begin{eqnarray}
\chi_t-\chi_{xx}-\chi=0.
\end{eqnarray}
Thus the transformation
\begin{eqnarray}
u=1-\frac{1}{1+\chi},
\end{eqnarray}
where $\chi$ satisfies the linear heat equation (5), is the 
linearizing transformation for eq.(1) in (1+1) dimensions for 
the choice $m=2$, which is similar to the one considered in reference [6]. Here we have given 
an interpretation for the transformation in terms of the B\"acklund 
transformation. Similar analysis holds good in the case of (2+1) 
dimensions of eq.(1) also and the same transformation (6) 
linearizes eq.(1) for $m=2$ as well, where $\chi$ satisfies the two dimensional linear heat equation $\chi_t-\chi_{xx}-\chi_{yy}-\chi=0$.

Now let us consider the invariance of eq.(1) under one-parameter
continuous Lie group of transformations. First we consider the 
(1+1) dimensional case,
\begin{eqnarray}
\displaystyle{u_t-u_{xx}-\frac{m}{1-u}u_x^2-u+u^2=0}.
\end{eqnarray}
Considering the infinitesimal transformation
\[
x \longrightarrow X = x+\varepsilon\xi(t,x,u), \;\;\;t \longrightarrow T
= t+\varepsilon\tau(t,x,u),
\]
\begin{eqnarray}
u \longrightarrow U = u+\varepsilon\phi(t,x,u),  \varepsilon <<1,
\end{eqnarray}
one can check that the invariance analysis of eq.(7) singles out the special value $m=2$ as the only choice having nontrivial Lie point symmetries of the form
\begin{equation}
 \tau  =  a, \quad \xi  =   b, \quad \phi  = c(t,x)(1-u)^2,
\end{equation} 
where $ a, b $ are arbitrary constants and  $ c(t,x) $
is any solution of the linear heat equation  
\begin{eqnarray}
c_t-c_{xx}-c =0.
\end{eqnarray}
The corresponding symmetry algebra is of the form
\begin{eqnarray}
[X_1,X_2]=0,\;[X_1,X_c]=X_{c_t}, \; [X_2,X_c]=X_{c_x},
\end{eqnarray}
where 
\begin{eqnarray}
X_1=\partial_t, \quad X_2 = \partial_x \;\mbox{and} \quad
X_c = c(t,x)(1-u)^2 \partial_u,
\end{eqnarray}
and is of infinite dimensional in nature.
However, for all other values of $m$ $(m\ne2)$ one obtains only the 
trivial translation symmetries 
\begin{equation}
\tau = a,\quad \xi = b,\quad \phi = 0. 
\end{equation} 
Solving now the characteristic equation associated with (9) for $m=2$,
\begin{eqnarray} 
\ds
{\frac {dt} {a} \quad = \quad \frac {dx}{b} \quad = \quad
\frac{du} {c(t,x)(1-u)^2}},
\end{eqnarray}
one obtains the similarity variables  
\begin{eqnarray} 
z=ax-bt,\quad  u=1- \ds {\frac {a}{a+\bar{w}(z)+\int c(t,x)dt}},  
\end{eqnarray} 
where $\bar{w}$ satisfies the similarity reduced ordinary differential equation (ode) of the form
\begin{eqnarray} 
\ds
{a^2 \bar {w}''+b\bar {w}' + \bar {w}=0}.
\end{eqnarray}
Eq.(16) is infact the form of the linear heat equation in terms of the 
wave variable $z$. Further, one can consider  the quantity
\begin{eqnarray} 
\ds {\frac {1}{a}\left [w(z)+\int c(t,x)dt\right ]}=\chi,  
\end{eqnarray}
where $\chi$ satisfies eq.(5), so that  the similarity variable 
$u$ in (15) is nothing but the linearizing transformation (6). It
has now been given a group theoretical interpretation.

Further, since the general solution of (16) is
\begin{eqnarray}
\bar {w} = I_1 e^{\ds {{m_1z}}}+ I_2 e^{\ds {{m_2z}}}, \quad 
m_{1,2}=\frac{-b \pm \sqrt {b^2-4a^2}}{2a^2},
\end{eqnarray}
where $I_1$ and $I_2$ are integration constants, we can write  
the solution to the original pde  as
\begin{eqnarray}
\fl u= \left \{
	\begin{array}{lll}
1-\ds{\frac {a}
{a+I_1 e^{\ds{{m_1(ax-bt)}}}+I_2 e^{\ds{{m_2(ax-bt)}}}+
\int c(t,x) dt}}, \qquad b^2-4a^2 > 0;
\\
1-\ds{\frac{a}{ a+e^{\ds{{p(ax-bt)}}} \left ( I_1+I_2(ax-bt) \right )
+\int c(t,x)dt }},\;\; \;\;\;\;\qquad b^2-4a^2 = 0; \\
1-\ds{\frac{a}{a+e^{\ds{{p(ax-bt)}}} \left ( I_1 \cos {q(ax-bt)} 
+ I_2 \sin {q(ax-bt)} \right )
+\int c(t,x)dt}},\\ \qquad  \mskip 420 mu b^2-4a^2 < 0
\end{array}
\right. 
\end{eqnarray}
with $p = -b/2a^2$,\; $q = \sqrt{4a^2-b^2}/2a^2$.
In the special case $c(t,x)=0$ one obtains all the interesting travelling 
wave solutions and stationary structures discussed in [7]. 

For all 
other values of 
$m$ $(\ne 2)$, one obtains from the symmetries (13) the similarity variables $z=ax-bt$ and $u=w(z)$ 
leading to an ode of the form
\begin{eqnarray} 
a^2 \bar {w} \bar {w}'' - ma^2 \bar {w}^{'2} + b \bar {w} \bar {w}' 
- (1- \bar {w}) \bar {w}^2=0 
\end{eqnarray} 
with $\bar {w} = 1-w$. The above eq.(20) is in general of 
non-Painlev\'e type when $a,b\ne 0$ except for $m=0$ and $b=5/\sqrt{6}$ under proper rescaling [9].
For the other choices one can solve for the case $b=0$ in terms of 
elliptic function solutions including the limiting case of solitary 
pulse which are all of static form. 

Now we extend our above analysis to the (2+1) dimensional case,
\begin{eqnarray}
\displaystyle{u_t-u_{xx}-u_{yy}-\frac{m}{1-u}(u_x^2+u_y^2)-u+u^2=0}.
\end{eqnarray}
Under the infinitesimal transformation
\[
\fl x \longrightarrow X = x+\varepsilon\xi(t,x,u),\;\;\;
y \longrightarrow Y = y+\varepsilon\eta(t,x,u),  \;\;\;t \longrightarrow T
= t+\varepsilon\tau(t,x,u),
\]
\begin{eqnarray}
\fl u \longrightarrow U = u+\varepsilon\phi(t,x,u),  \varepsilon <<1,
\end{eqnarray}
the invariance analysis separates out the special value of $m=2$
for which eq.(21) possesses the following Lie point symmetries:
\begin{eqnarray}
\tau=a,
\quad \xi=b_3y+b_4, \quad \eta = -b_3x+d_4,\quad \phi = c(t,x,y)(1-u)^2 
\end{eqnarray}
where $b_3,\;b_4$ and $d_4$ are arbitrary constants and 
$ c(t,x,y) $ is the solution of the two 
dimensional linear heat 
equation
\begin{eqnarray}
c_t-c_{xx}-c_{yy}-c = 0. 
\end{eqnarray}
But for all other choices of $m$ $(\ne 2)$ we get 
\begin{eqnarray} 
\tau=a, \quad
\xi=b_3y+b_4, \quad \eta = -b_3x+d_4, \quad \phi = 0.
\end{eqnarray} 

As earlier, for $m=2$, the similarity variables are found by solving the
characteristic equation associated with the symmetries (23). They
are 
\begin{eqnarray} 
\fl & \ds {z_1=\frac{b_3}{2}(x^2+y^2)+b_4y-d_4x,\quad z_2=-t
-\frac{a}{b_3}\sin^{-1}\left(\frac{d_4-b_3x}{\sqrt{d_4^2+2b_3z_1+b_4^2}}
\right),} & \nonumber\\
\fl  &  u=1- \ds { \frac {a}{w(z_1,z_2)+\int c(t,x,y)dt}}.&
\end{eqnarray}
where $w$ satisfies the similarity reduced (1+1) dimensional pde
\begin{eqnarray}
\fl \ds {w_{z_2}+2b_3w_{z_1}+(2b_3z_1+b_4^2+d_4^2)w_{z_1z_1}+\frac {a^2w_
{z_2z_2}}{2b_3z_1+b_4^2+d_4^2}+w-a=0}.
\end{eqnarray}
The above equation is nothing but the linear heat equation in terms of the variables $z_1$ and $z_2$. Here too as in eq.(15) one can 
consider the similarity form  (26) in two dimensions to get the linear heat equation
\begin{eqnarray}
\chi_t-\chi_{xx}-\chi_{yy}-\chi=0,\quad \chi=\frac{1}{a}\left [w(z_1,z_2)+\int c(t,x,y)dt\right ], 
\end{eqnarray}
so that the transformation for the variable $u$ can again be 
 interpreted as the linearizing transformation as in the 1-dimensional case from a group theoretical point of view.

Carrying out again a Lie symmetry analysis on eq.(27),  one  obtains  the 
similarity variables 
\begin{eqnarray} 
\fl &\zeta=\bar {z}_1,
\quad {w}= a+e^{ \left (\ds{ \frac {c_1\bar {z}_2}{c_3}}\right
)}\left [f(\zeta)+\ds{\frac{1}{c_3} \int c_2(\bar {z}_1, \bar {z}_2)} e^{\ds{ \left
(-{\frac {c_1}{c_3}} \bar {z}_2 \right )}} d\bar {z}_2  
\right ],\nonumber&
\\ 
\fl &\bar {z_1} = 2b_3z_1+b_4^2+d_4^2, \quad
\bar {z}_2 = z_2, \quad b_3,d_4\ne 0, &
\end{eqnarray} 
where $f$ satisfies the linear second order ode of the form 
\begin{eqnarray} 
\fl \zeta^2 f''+\zeta f'+(A+B \zeta)f=0,\quad A=
(ac_1/2b_3c_3)^2 \quad B = (1+c_1/c_3)/4b_3^2
\end{eqnarray} 
with prime  denoting differentiation w.r.t. $
\zeta $. 

As
the general solution to eq.(30) can be expressed in terms of
cylindrical functions of the form 
\begin{eqnarray} 
f=I_1Z_1(2\sqrt{B\zeta})+I_2Z_2(2\sqrt{B\zeta}), 
\end{eqnarray} 
where $ Z_i$'s, $ i=1,2 $, are the two
linearly independent cylindrical functions and $I_1$ and $I_2$ are
arbitrary constants, the invariant solution to the 
(2+1)-dimensional pde
(21) is written as
\begin{eqnarray} 
\fl u & = & 1- a \Bigg [ a + e^{\ds {\left(
{\frac{c_1}{c_3}} \bar{z}_2\right )}} \Bigg (I_1Z_1(2\sqrt{B\bar{z}_1})
+ I_2Z_2(2\sqrt{B\bar{z}_1}) \nonumber \\  
\fl &&  - \ds{\frac{1}{c_3}\int c_2(\bar{z}_1,\bar{z}_2)}
e^{\ds{\left({\frac{c_1}{c_3}\bar{z}_2}\right)}} 
d\bar{z}_2\Bigg) 
 +\ds{\int c(t,x,y)dt} \Bigg]^{-1}.
\end{eqnarray}
Next, proceeding to the special case, $ b_3 =0 $ and $ d_4 =0 $ 
in eq.(26), and carrying out an analysis as above,  we 
get the solution to the original pde (21) as
\begin{eqnarray}
\fl u=\left \{
	\begin{array}{lll}
1-a\Bigg \{a+\exp\Bigg [\ds{{-k\left (\frac{a}{b_4}x-t\right )}}\Bigg ]\Bigg [ I_1\cos(\sqrt{k_1}c_5b_4y)+I_2\sin(\sqrt{k_1}c_5b_4y) \Bigg . \Bigg .\\ 
\Bigg. \Bigg.
\mskip 150 mu+\ds{\int \frac{c_3(z_1,z_2)}{c_5}}e^{\ds {kz_2}}dz_2 \Bigg ] 
+\ds {\int c(t,x,y)dt}\Bigg \}^{-1}, \; \; k_1<0, \\ [2mm]
1-a\Bigg \{a+\exp\Bigg [\ds {-k\left (\frac{a}{b_4}x-t \right )}\Bigg ] \Bigg [ I_1e^{\ds{\sqrt{k_1}c_5b_4y}}+I_2e^{\ds{-\sqrt{k_1}c_5b_4y}}\Bigg . \Bigg. \\
\Bigg. \Bigg. 
\mskip 150 mu +\ds{\int \frac{c_3(z_1,z_2)}{c_5}}e^{\ds{kz_2}}dz_2 \Bigg ]
+\ds{\int c(t,x,y)dt}\Bigg \}^{-1}, \; \; k_1>0 \\[2mm]
1-a\Bigg \{a+\exp\Bigg[\ds{-k\left (\frac{a}{b_4}x-t \right )}\Bigg ] \Bigg [ I_1c_5b_4y+I_2 \Bigg. \Bigg. \\ \Bigg. \Bigg. \mskip 150 mu
+\ds{\int \frac{c_3(z_1,z_2)}{c_5}}e^{\ds{kz_2}}dz_2\Bigg ]
+\ds{\int c(t,x,y)dt} \Bigg \}^{-1}, \; \; k_1=0 \\[2mm]
\end{array}
\right. 
\end{eqnarray}
where the parameter $k_1=\ds{\frac{1}{b_4^2c_5^2}}\left[\ds{k}-\left(\ds{\frac{ak}{b_4}}\right)^2-1\right]$ 
with  $k=-c_2/c_5$ and $z_1=b_4y, \; z_2 = \ds{\frac{a}{b_4}}x-t$. Here $c_2,\;c_4,\;c_5$ are arbitrary constants of
integration.

Eqs.(32) and (33) contain a large class of interesting solutions 
of various types including travelling wave solutions and static 
patterns. Particularly, we can easily check that all the 
five classes of travelling wave solutions 
discussed in [7] can be derived from eq.(33) for particular choices of the 
constants involved along with the specific choice that the 
functions  $c_3(z_1,z_2) = 0$ and $c(t,x,y) = 0$. 
Specifically, the  simplest {\emph{travelling wave solution}} 
\begin{eqnarray}
u=1-\ds{\frac{1}{1 + A \exp \left [-k\left (\ds{\frac{a}{b_4}x-t}\right )\pm\sqrt{k_1}c_5b_4y \right  ]}}, \; \; k_1>0
\end{eqnarray}
can be constructed by assuming either $I_1 = 0$ or $ I_2 = 0$.
However, if we  choose 
$I_1=I_2 \; (\ne 0)$ in eq.(33), we obtain the {\emph{V-wave}} solution 
\begin{eqnarray}
u=1-\ds{\frac{1}{1 + A \exp \left [\ds{-k\left (\frac{a}{b_4}x-t\right )}\right ]\cosh(\sqrt{k_1}c_5b_4y)}}, \; \; k_1>0.
\end{eqnarray}
On the other hand   for $I_2=0$ and $k_1 < 0 $ in eq. (33), we
obtain  the \emph{ wave front oscillating} in space:
\begin{eqnarray}
u=1-\ds{\frac{1}{1 + A \exp \left[\ds{-k\left (\frac{a}{b_4}x-t\right )}\right ]\cos(\sqrt{k_1}c_5b_4y)}}. 
\end{eqnarray}
The case $k_1=0$, $I_1=0$ describes a travelling plane wave. But 
when $I_1\ne 0$ and $I_2=0$ we get an inhomogeneous solution 
\begin{eqnarray}
u=1-\ds{\frac{1}{1 + A |y|\exp \left [\ds{-k\left (\frac{a}{b_4}x-t\right )}\right ]}}
\end{eqnarray}
which is nothing but the \emph{separatrix} solution. Finally one 
more choice exists for positive $k_1$ and $ I_1=-I_2 $:
\begin{eqnarray}
u=1-\ds{\frac{1}{1 + A \exp\left [\ds{-k\left(\frac{a}{b_4}x-t\right)}\right ]|\sinh(\sqrt{k_1}c_5b_4y)|}}. 
\end{eqnarray}
This  is the \emph{Y-wave} solution. In each of  the above solutions $A$ is a positive constant. Several static structures can also be obtained as limiting cases of the above solutions (32) and (33).

Proceeding in a similar fashion for $m$ other than $2$ in eq.(21), we obtain  the similarity reduced variables
\begin{eqnarray}
\fl & \ds {z_1=\frac{b_3}{2}(x^2+y^2)+b_4y-d_4x,\quad z_2=-t
-\frac{a}{b_3}\sin^{-1}\left(\frac{d_4-b_3x}{\sqrt{d_4^2+2b_3z_1+b_4^2}}
\right), }& \nonumber\\
\fl &  u=w(z_1,z_2), \quad (b_3,d_4 \ne 0). &
\end{eqnarray}
along with the reduced pde of the form 
\begin{eqnarray}
\ds{w_{z_2}+2b_3w_{z_1}+(2b_3z_1+b_4^2+d_4^2)w_{z_1z_1}+\frac{a^2
w_{z_2z_2}}
{2b_3z_1+b_4^2+d_4^2}} +\frac {m}{1-w} \nonumber\\
\ds{\left[(2b_3z_1+b_4^2
+d_4^2)w_{z_1}^2+\frac{a^2w_{z_2}^2}{2b_3z_1+b_4^2+d_4^2}\right]
+w-w^2=0}. 
\end{eqnarray} 
On carrying out a similarity reduction, eq.(40) reduces to an ode 
\begin{eqnarray}
4b_3^2\left [\zeta f''+ \left(f'+\frac
{m\zeta}{1-f}f^{'2}\right)\right ]+f-f^2=0, \quad '=d/d\zeta
\end{eqnarray}
where 
\begin{eqnarray}
\zeta=\bar{z}_1, \quad w=f(\zeta),\quad \bar{z}_1=2b_3z_1+b_4^2+d_4^2,
\end{eqnarray}
giving rise to static structures in $(x,y)$ variables.
This equation is found to be nonintegrable, in general. However, if we consider  for the special
case $b_3=0, \; d_4=0$, the similarity variables  become $z_1=b_4y,\; z_2=\ds{\frac{a}{b_4}x-t}, \; u =w(z_1,z_2)$, eq.(40) on one more 
reduction gives  the ode 
\begin{eqnarray}
Df''+\frac{Dm}{1-f}{f'}^2-c_1f'+f(1-f)=0, \nonumber \\ D=\left (\frac{a^2}{b_4^2}{c_1}^2+b_4^2{c_2}^2\right ), \quad '=d/d\zeta.
\end{eqnarray}
with $\zeta=-c_1(\frac{a}{b_4}x-t)+c_2b_4y \; \mbox{and}\; w = f(\zeta)$, giving rise to plane wave solutions. 
This equation is of non-Painlev\'e type, except for the choice $m=0$ in which case  eq.(43) reduces to the form (20) with $m=0$.

On the other hand with the choice $b_3=0$ alone, the similarity variables become $ z_1=d_4x-b_4y,\; z_2=ax-b_4t \; \mbox{and}\;  u=w(z_1,z_2)$. 
 Correspondingly eq.(40) on a  further reduction reduces to an ode 
\begin{eqnarray} 
Af_1''+Bf_1'-\frac{Am}{f_1}f_1^{'2}-f_1+f_1^2=0, \; '=d/d\zeta
\end{eqnarray} 
with $ f_1=1-f, \;\;A=a^2(c_1^2d_4^2+c_2^2b_4^2),\; 
B=-d_4b_4(c_1+c_2)\; \mbox{and}\; \zeta = ac_2(d_4x-b_4y)-d_4(c_1+c_2)(ax-b_4t),\; w=f(\zeta)$. However, the system is found to possess  elliptic function solutions 
including the limiting case of the solitary pulse for particular choices of the constants involved.  More details of these results 
will be published elsewhere. 

Throughout our above analysis we have made use of the computer program MUMATH
[10] to determine the symmetries.

To conclude, the generalized Fisher type equation with modified diffusion in (1+1) and (2+1) dimensions has been found to be integrable 
only for the special case $m=2$ via both the singularity structure and symmetry analysis. Moreover we have also shown that a B\"acklund transformation gives rise to the linearizing transformation to the heat equation for the integrable case. Besides, a Lie symmetry analysis leads in a natural way to the linearizing transformation and physically important classes of solutions through  similarity variables and similarity reductions, thereby giving a group theoretical understanding of the system.

\ack

This work forms a part of the National Board of Higher Mathematics,
Department of Atomic Energy, Government of India and the Department 
of
Science and Technology, Government of India research projects. M. S. 
wishes to thank the University of Sydney for providing a Post 
Doctoral Fellowship.

\end{document}